# Optical phase extraction algorithm based on the continuous wavelet and the Hilbert transforms

Mustapha Bahich, Mohamed Afifi, Elmostafa Barj

**Abstract**— In this paper we present an algorithm for optical phase evaluation based on the wavelet transform technique. The main advantage of this method is that it requires only one fringe pattern. This algorithm is based on the use of a second π/2 phase shifted fringe pattern where it is calculated via the Hilbert transform. To test its validity, the algorithm was used to demodulate a simulated fringe pattern giving the phase distribution with a good accuracy.

——————————— ◆ ———————————

## 1. INTRODUCTION

TODAY, optical techniques have been employed in many sciences and engineering applications to compute several physical magnitudes which are codified as the phase of a periodic intensity profiles, so the development of more sophisticated phase extraction algorithms is continuously needed [1,2].

Two classical methods for phase extraction are phase-shifting technique [3] and Fourier transform technique [4]. Phase-shifting technique processes the fringe patterns pixel by pixel. Each pixel is processed separately and does not influence the others. However, three, four or more images are needed. On the contrary, Fourier transform technique processes the whole frame of a fringe pattern at the same time, but it requires a spatial carrier and the pixels will influence each other. Thus a compromise between the pixel-wise processing and global processing is necessary. One of the solutions is to process the fringe patterns locally. For this, different wavelet algorithms are conceived to extract the phase distribution of the fringe patterns. The wavelet concept and its applications, is becoming a useful tool in various studies for analyzing localized variations and particularly to analyze non-stationary or transient signals [5].

The aim of this paper is the application of the continuous wavelet analysis to extract the optical phase distribution from a single recorded fringe pattern without high frequency spatial carrier. The method applied requires a second phase shifted fringe pattern which will be generated numerically via the Hilbert Transform [6]. The use of a single image can lead to the phase distribution of dynamic processes. It seems suitable where only one fringe pattern can be taken and it is also applicable on the fringe patterns without spatial carrier [7].

The paper first presents an introduction of the continuous wavelet transform. In section 3, we expose the wavelet phase extraction method. In sections 4 and 5, we present the mathematical description of the corrected phase shift via Hilbert Transform. Finally, results of numerical simulations are presented in section 6.

## 2. THE CONTINUOUS WAVELET TRANSFORM

The continuous wavelet transform (CWT) is a powerful tool to obtain a space–frequency description of a signal. Unlike the Fourier transform that uses an infinitely oscillating terms $e_\omega = \exp(-i\omega x)$, the wavelet analysis technique use a set of a specially designed pulse functions, called "wavelets", to analyze the local information of the signal[8]. A wavelet is an oscillating function $\psi(x)$, centered at $x=0$ and decay to zero such

$$\int_{-\infty}^{+\infty} \psi(x)dx = 0 \qquad (1)$$

If $\hat{\psi}(\omega)$ is the Fourier transform of $\psi(x)$, then condition (1) is equivalent to the requirement that

$$\hat{\psi}(0) = 0 \qquad (2)$$

A family of the analyzing wavelets is generated from this "mother wavelet" $\psi(x)$ by translations and dilations and it can be expressed as

$$\psi_{s,\xi}(x) = \frac{1}{\sqrt{s}} \psi\left(\frac{x-\xi}{s}\right) \qquad (3)$$

Where $s \neq 0$ is the scale parameter related to the frequency concept, and $\xi \in R$ is the shift parameter related to position. We note that the wavelets with small values of s have narrow spatial support and consequently rapid oscillations, making them well adapted to selecting high-frequency components of a signal. The converse is true for wavelets with large values of s.

Many different types of mother wavelets are available for the phase evaluation applications, but in our case the study reveals that the Complex Morlet wavelet gives the best results. It is defined as

$$\psi(x) = \frac{1}{\sqrt{\pi}} \exp(2i\pi f_c x)\exp(-x^2) \qquad (4)$$

where $f_c$ is the wavelet center frequency.
Fig.1 shows the real and the imaginary part of the Complex Morlet wavelet.

---

- M. Bahich is with Applied optics and image processing group, Faculty of Sciences Ben Msik, Hassan II-Mohammedia University, Morocco.
- M. Afifi is with the Applied optics and image processing group, Faculty of Sciences Ben Msik, Hassan II-Mohammedia University, Morocco.
- E. M. Barj is with the Applied optics and image processing group, Faculty of Sciences Ben Msik, Hassan II-Mohammedia University, Morocco.

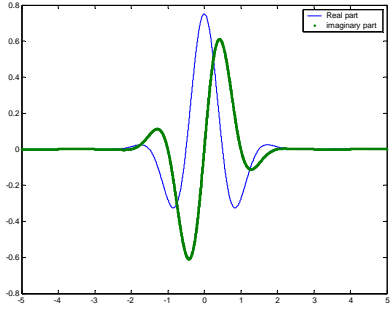

Fig. 1. Complex Morlet wavelet

The 1D continuous wavelet transform (CWT) of a function $f(x)$ is giving by

$$W_f(s,\xi) = \langle f, \psi_{s,\xi} \rangle = \int_{-\infty}^{+\infty} f(x) \psi_{s,\xi}^*(x) dx \quad (5)$$

where * denotes the complex conjugation.

The continuous wavelet transform can be expressed, using the Parseval identity, as

$$W_f(s,\xi) = \frac{1}{2\pi} \langle \hat{f}, \hat{\Psi}_{s,\xi} \rangle$$
$$= \frac{\sqrt{s}}{2\pi} \int_{-\infty}^{+\infty} \hat{f}(k) \left(\hat{\Psi}(sk)\right)^* e^{i\xi k} dk \quad (6)$$

Where $\hat{f}$ and $\hat{\psi}$ and $k$ are respectively the Fourier transform of the signal, the Fourier transform of the mother wavelet and the angular frequency.

If the inverse wavelet transform exist, the original signal can be reconstructed by

$$f(x) = \frac{1}{C_\Psi} \int_0^{+\infty} \int_{-\infty}^{+\infty} W_f(s,\xi) \Psi_{s,\xi}(x) \frac{ds\,d\xi}{s^2} \quad (7)$$

where

$$C_\Psi = \int_{-\infty}^{+\infty} \frac{|\hat{\Psi}(k)|^2}{k} dk \quad (8)$$

This reconstruction of the signal is possible when $C_\Psi$ has a finite value.

## 3. WAVELET PHASE EXTRACTION METHOD

There are many techniques for extracting phase distributions from two-dimensional fringe patterns.

The fringe patterns, derived from two-beam interferometers, can be mathematically formulated by the sinusoidal dependence of the intensity on the spatial coordinates $(x,y)$ of the image plane:

$$I(x,y) = I_0(x,y)\left[1 + V(x,y)\cos(\phi(x,y))\right] \quad (9)$$

Where $I_0$ is the bias intensity, $V$ the visibility or fringe contrast and $\phi$ the optical phase.

The one-dimensional continuous wavelet transform of the fringe pattern's row $x$ (in the $y$ direction) is given by

$$W(x,s,\xi) = \frac{1}{\sqrt{s}} \int_{-\infty}^{+\infty} I(x,y) \left(\Psi\left(\frac{y-\xi}{s}\right)\right)^* dy \quad (10)$$

Exploiting the wavelet localization property and assuming a slow variation of the intensity bias and the visibility, the wavelet transform becomes

$$W(x,s,\xi) = \frac{I_0(x,\xi) V(x,\xi)}{\sqrt{s}}$$
$$\cdot \int_{-\infty}^{+\infty} \cos\left[\phi(x,\xi) + (y-\xi)\frac{\partial\phi}{\partial y}(x,\xi)\right] \left(\Psi\left(\frac{y-\xi}{s}\right)\right)^* dy \quad (11)$$

and the Parseval identity leads to

$$W(x,s,\xi) = \frac{I_0(x,\xi) V(x,\xi) \sqrt{s}}{2} \int_{-\infty}^{+\infty} h(x,k) \left(\hat{\Psi}(sk)\right)^* e^{i\xi k} dk \quad (12)$$

with

$$h(x,k) = \delta\left(k - \frac{\partial\phi}{\partial y}(x,\xi)\right) \exp\left[i\left(\phi(x,\xi) - \xi \frac{\partial\phi}{\partial y}(x,\xi)\right)\right]$$
$$+ \delta\left(k + \frac{\partial\phi}{\partial y}(x,\xi)\right) \exp\left[-i\left(\phi(x,\xi) - \xi \frac{\partial\phi}{\partial y}(x,\xi)\right)\right] \quad (13)$$

Finally, the wavelet transform becomes

$$W(x,s,\xi) = \frac{I_0(x,\xi) V(x,\xi) \sqrt{s}}{2} \left[\left(\hat{\Psi}\left(s\frac{\partial\phi}{\partial y}\right)\right)^* e^{i\phi(x,\xi)}\right.$$
$$\left. + \left(\hat{\Psi}\left(-s\frac{\partial\phi}{\partial y}\right)\right)^* e^{-i\phi(x,\xi)}\right] \quad (14)$$

The two terms in the previous equation in general overlap and have to be separated in order to retrieve the phase. The method commonly used is to add a carrier frequency m to the signal satisfying [9, 10]:

$$m > \left|\frac{\partial\phi}{\partial y}\right|_{max} \quad (15)$$

In this study we use an alternative method based on the use of a second π/2 phase shifted fringe pattern expressed by

$$I_d(x,y) = I_0(x,y)\left[1 + V(x,y)\cos\left(\phi(x,y) + \frac{\pi}{2}\right)\right]$$
$$= I_0(x,y)\left[1 - V(x,y)\sin(\phi(x,y))\right] \quad (16)$$

And its CWT is

$$W_d(x,s,\xi) = i\frac{I_0(x,\xi) V(x,\xi) \sqrt{s}}{2}\left[-\left(\hat{\Psi}\left(s\frac{\partial\phi}{\partial y}(x,\xi)\right)\right)^* e^{i(\phi(x,\xi))}\right.$$
$$\left. + \left(\hat{\Psi}\left(-s\frac{\partial\phi}{\partial y}(x,\xi)\right)\right)^* e^{-i(\phi(x,\xi))}\right] \quad (17)$$

Then using (14) and (17) we get

$$W_s(x,s,\xi) = W(x,s,\xi) + i W_d(x,s,\xi) \quad (18)$$

$$W_s(x,s,\xi) = I_0(x,\xi) V(x,\xi) \sqrt{s}\left(\hat{\Psi}\left(s\frac{\partial\phi}{\partial y}(x,\xi)\right)\right)^* e^{i(\phi(x,\xi))} \quad (19)$$

The wavelet coefficients $W_s(x,s,\xi)$ of the row $x$ is a matrix which its modulus and phase arrays can be calculated by the following equations:

$$abs(s,\xi) = |W_s(x,s,\xi)| \quad (20)$$

$$\phi(s,\xi) = \tan^{-1}\left(\frac{Im\{W_s(x,s,\xi)\}}{Re\{W_s(x,s,\xi)\}}\right) \quad (21)$$

To compute the phase distribution of the row $x$, the maximum value of each column of the modulus array is determined and then its corresponding phase value is found from the phase array. By repeating this process to all rows of the fringe pattern, the wrapped phase map is resulted. However, the phase resultant is modulo π and needs to be extended to modulo 2π before unwrapping it [11].

## 4. THE HILBERT TRANSFORM

To extract the phase distribution from a single fringe pattern without spatial carrier, the π/2 phase shifted fringe pattern Id must be generated numerically by Hilbert transform.
In the 1D real space, the Hilbert transform (HT) of the signal $f(x)$ is a convolution between it and $\frac{1}{\pi x}$ [12]. It is defined by

$$H(f(x)) = \frac{1}{\pi} \int_{-\infty}^{+\infty} \frac{f(x')}{(x-x')} dx' \quad (22)$$

In the frequency domain, the Hilbert transform results from a simple multiplication with the sign function:

$$FT(H(f))(k) = -i\,\text{sgn}(k)FT(f)(k) \quad (23)$$

where $FT$ denotes the Fourier transform, $k$ is the angular frequency and

$$\text{sgn}(k) = \begin{cases} -1 & \text{for } k < 0 \\ 0 & \text{for } k = 0 \\ 1 & \text{for } k > 0 \end{cases} \quad (24)$$

Hence the Hilbert transform is equivalent to a filter altering the phases of the frequency components by 90° positively or negatively according to the sign of frequency.
the row of the fringe pattern is given by:

$$I(x) = I_0(x) + M(x)\cos(\phi(x)) \quad (25)$$

Where

$$M(x) = I_0(x)V(x) \quad (26)$$

Assuming a slow variation of the background intensity, $I_0$ must be removed by an appropriate high-pass filter before the application of HT, then we obtain

$$I_f(x) = M(x)\cos(\phi(x)) \quad (27)$$

When $M(x)$ doesn't have a frequency's components beyond the $\frac{\partial \phi}{\partial x}$, then Hilbert transform of $I_f(x)$ is [13]:

$$I_d(x) = -M(x)\sin(\phi(x)) \quad (28)$$

## 5. SIGN CORRECTION

Admittedly, the Hilbert transform gives the desired shifted interferogram when the phase is monotonically increasing, but in the other case, applying the Hilbert transform to an interferogram with closed fringes leads to a π/2 phase shifted fringe pattern containing a sign ambiguity. To obtain the phase exact shifted interferogram, we need to know the sign of the local spatial frequencies to correct this ambiguity. For that, we propose to compute the fringes orientation angle to get the sign map of the local spatial frequencies [14].
Orientation estimation plays an important role in image processing and image coding, for texture analysis, adaptive filtering and image enhancement. Fringe orientation provides important information for analyzing and understanding fringes.

The orientation angle of the interferogram pixel with the coordinates (x,y) is:

$$\theta(x,y) = arctg\left(\frac{\partial \phi(x,y)}{\partial y} \bigg/ \frac{\partial \phi(x,y)}{\partial x}\right) \pm \frac{\pi}{2} \quad (29)$$

However, the phase gradients are not directly accessible, so we have to use the acquired interferogram to estimate it. In this work, we use the gradient-based method to compute the fringe orientation, due to its accuracy and computation time [15]. Fig.2 is an illustration of a fringe orientation angle.

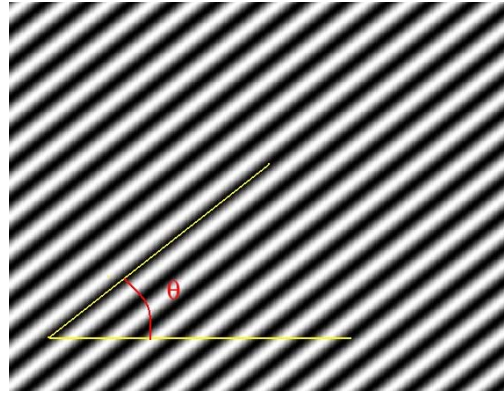

Fig. 2. Orientation angle θ for a texture with a cosine profile function.

The fringe orientation angle that we get is modulo π and should be modulo 2π. This can be achieved by unwrapping process. The unwrapped version of the orientation corresponds to the present term "direction angle".

## 6. SIMULATION RESULTS

In our numerical simulation, we used the Complex Morlet as a mother wavelet to verify the ability of the method to determinate the phase distribution with a quite good accuracy.
We generate a fringe pattern of size 512 × 512 which is shown in the Fig.3.

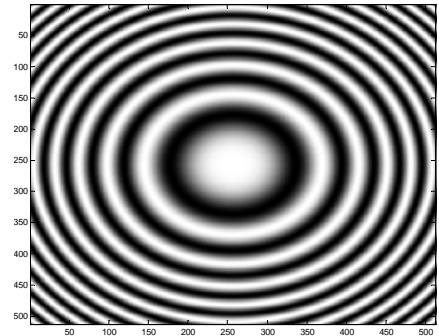

Fig. 3. The simulated fringe pattern

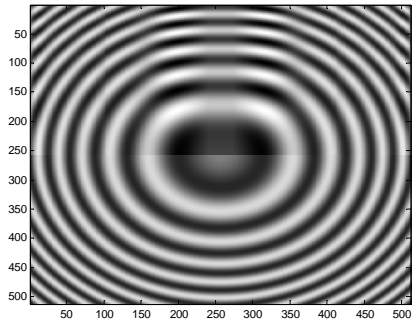

Fig. 4. The π/2 phase shifted fringe pattern: The estimated fringe pattern (Upper part) and the theoretical fringe pattern (Lower part)

The intensity distribution of this fringe pattern is given by
$$I(x, y) = 1 + 0.5\cos(\phi(x, y)) \qquad (30)$$
The resulting fringe pattern of the corrected Hilbert Transform is illustrated in the Fig.4.

The test phase function $\phi(x, y)$ that we used has the following expression:
$$\phi(x, y) = 0.0005\left((x - 256)^2 + (y - 256)^2\right) \qquad (31)$$

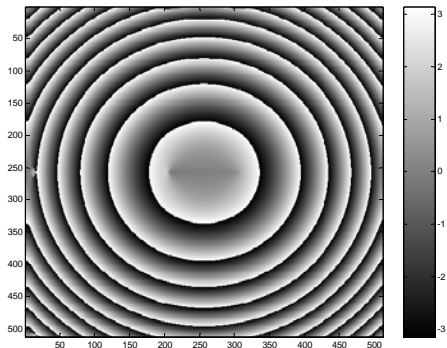

Fig. 5. Wrapped retrieved phase.

The wavelet phase extraction algorithm is applied to the two combined fringe patterns of Fig.3 and Fig.4, to retrieve the phase distribution. The wrapped recovered phase distribution is presented in the Fig.5.
Fig.6.a and Fig.6.b show the theoretical phase and the unwrapped retrieved phase distributions respectively.

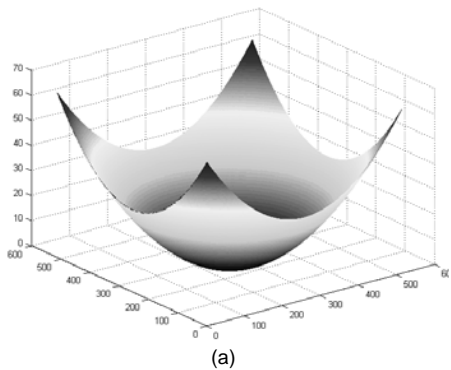

(a)

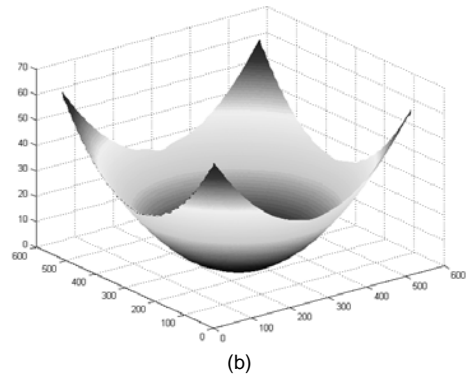

(b)

Fig. 6. (a) Exact phase distribution, (b) Extracted phase distribution.

## 7. CONCLUSION

In this work, we prove the possibility to extract the optical phase from a single interferogram with open or closed fringes. In this technique, the continuous wavelet transform is applied on two phase quadrature interferograms for estimating the phase distribution. The Hilbert transform with the amplitude and phase correction are utilized for the purpose of the second π/2 phase shifted interferogram estimation.

Consequently, this technique seeks to extract the phase of the interferogram without the need for special carrier, which make it applicable to the study of transient phenomenas. The performance of this technique was evaluated through a simulated example.

## REFERENCES

[1] R. S. Sirohi and F. S. Chau, Optical Methods of Measurement, Marcel Dekker Inc., 1999.

[2] B. V. Dorrio, and L. L. Fernandez, "Phase-evaluation methods in whole-field optical measurement techniques", Meas. Sci. Technol. 10 (1999), 33-55.

[3] Creath K. Phase-measurement interferometry techniques. In: Wolf E, editor. Progress in optics, vol. 26. Amsterdam: Elsevier; 1988. p. 349–93.

[4] Takeda M, Ina H, Kobayashi S. Fourier transform methods of fringe-pattern analysis for computerbased topography and interferometry. J Opt Soc Am 1982; 72:156–60.

[5] I. Daubechies; "The wavelet transform time-frequency localization and signal analysis" IEEE Trans. Inf. Theory 36(1990), 961-1005.

[6] D. A. Zweig and R. E. Hufnagel, "A Hilbert transform algorithm for fringe-pattern analysis," Proc. SPIE Adv. Opt. Manuf. Testing, pp. 295–302, 1990.

[7] Tay C.J, Quan C, Yang F. J, He X.Y, A new method for phase extraction from a single fringe pattern, Optics communications 2004, vol. 239, no4-6, pp. 251-258

[8] G. Kaiser; "A Friendly Guide to Wavelets"; (Birkhauser, Boston, Mass., 1994).

[9] Afifi, M., Fassi-Fihri, M., Marjane, M., Nassim, K., Sidki, M. and Rachafi, S. Paul wavelet-based algorithm for optical phase distribution evaluation. Opt. Comm. 2002; 211, pp. 47-51.

[10] Qian Kemao, Seah H. S. and A. Asundi," Wavelets in optical metrology", in Perspectives in Engineering Optics, K. Singh, V. K. Rastogi eds., 117-133, Anita Publications, 2003.

[11] Ghiglia, D.C., and Pritt, M.D: 'Two-dimensional phase unwrapping: theory, algorithms and software' (Wiley, New York, 1998).

[12] Saff E. B., Snider A. D., Complex analysis for mathematics, science

and engineering, Prentice- Hall, Inc., New York, 1976.

[13] Bedrosian, E., 1963: A product theorem for Hilbert transform. Proc. IEEE, 51, 868–869.

[14] Kieran G. Larkin, "Natural demodulation of two-dimensional fringe patterns. II. Stationary phase analysis of the spiral phase quadrature transform," J. Opt. Soc. Am. A 18, 1871-1881 (2001)

[15] Xia Yang, Qifeng Yu and Sihua Fu, "A combined method for obtaining fringe orientations of ESPI " , Opt. Comm. Vol 273, 60-66 (2007)

**Mustapha Bahich** obtained his bachelor degree in applied physics followed by master in information processing techniques from Hassan II-Mohammedia University, Casablanca, Morocco. Currently, he is PhD candidate in Department of physics at Faculty of sciences Ben Msik, Hassan II-Mohammedia University, Casablanca, Morocco.

His current research interests include signal and image processing, fringe pattern analysis, phase unwrapping, wavelet techniques.